\documentclass[conference]{IEEEtran}
\usepackage[utf8]{inputenc}
\usepackage[T1]{fontenc}
\usepackage{microtype}
\usepackage{graphicx}
\usepackage{dblfloatfix}
\usepackage{filecontents}
\usepackage[noadjust]{cite}
\usepackage[hidelinks]{hyperref}
\usepackage{balance}
\usepackage{amsmath}
\usepackage{url}
\usepackage{csquotes}
\usepackage{mdframed}
\usepackage{subcaption}
\usepackage{multirow}
\usepackage[table]{xcolor}
\usepackage{enumitem}
\usepackage{bm}

\makeatletter
\newcommand{\textlabel}[2]{%
	\protected@edef\@currentlabel{#1}
	\phantomsection
	\label{#2}
}
\makeatother

\begin{document}

\title{The Potential of Using Vision Videos for CrowdRE: Video Comments as a Source of Feedback}

\author{
	\IEEEauthorblockN{
		Oliver Karras
	}
	\IEEEauthorblockA{
		TIB -- Leibniz Information Centre for\\Science and Technology\\
		Hannover, Germany\\ 
		Email: oliver.karras@tib.eu
	}
	\and
	\IEEEauthorblockN{
		Eklekta Kristo
	}
	\IEEEauthorblockA{ 
		Leibniz University Hannover\\
		Software Engineering Group\\
		Hannover, Germany\\ 
		Email: eklekta.kristo@stud.uni-hannover.de
	}
	\and
	\IEEEauthorblockN{
		Jil Klünder
	}
	\IEEEauthorblockA{
		Leibniz University Hannover\\
		Software Engineering Group\\
		Hannover, Germany\\ 
		Email: jil.kluender@inf.uni-hannover.de
	}
}

\maketitle 

\begin{abstract}
Vision videos are established for soliciting feedback and stimulating discussions in requirements engineering (RE) practices such as focus groups. Different researchers motivated the transfer of these benefits into crowd-based RE (CrowdRE) by using vision videos on social media platforms. So far, however, little research explored the potential of using vision videos for CrowdRE in detail. 
In this paper, we analyze and assess this potential, in particular, focusing on video comments as a source of feedback.
In a case study, we analyzed 4505 comments on a vision video from YouTube. We found that the video solicited 2770 comments from 2660 viewers in four days. This is more than 50\% of all comments the video received in four years. Even though only a certain fraction of these comments are relevant to RE, the relevant comments address typical intentions and topics of user feedback, such as \textit{feature request} or \textit{problem report}. Besides the typical user feedback categories, we found more than 300 comments that address the topic \textit{safety} which has not appeared in previous analyses of user feedback. In an automated analysis, we compared the performance of three machine learning algorithms on classifying the video comments. Despite certain differences, the algorithms classified the video comments well.
Based on these findings, we conclude that the use of vision videos for CrowdRE has a large potential. Despite the preliminary nature of the case study, we are optimistic that vision videos can motivate stakeholders to actively participate in a crowd and solicit numerous of video comments as a valuable source of feedback.
\end{abstract}


\IEEEpeerreviewmaketitle

\section{Introduction}
In the recent years, several researchers \cite{Bennaceur.2016, Darby.2018, Schneider.2019, Busch.2020, Vistisen.2017} introduced vision videos into Requirements Engineering (RE) for soliciting feedback and stimulating discussions. According to Karras~et~al.~\cite[p.~2]{Karras.2020a}, \enquote{\textit{a vision video is a video that represents a vision or parts of it} [(problem, solution, improvement)] \textit{for achieving a shared understanding among all parties involved by disclosing, discussing, and aligning their mental models of the future system}}. These videos are used to elicit and validate an integrated view of the future system regarding new and diverging aspects among all parties involved~\cite{Karras.2021a}.

In 2017, Schneider et al. \cite{Schneider.2017} proposed the use of vision videos for CrowdRE to solicit feedback. Videos are a well-suited medium for CrowdRE since their communication richness and effectiveness \cite{Ambler.2002, Karras.2018, Karras.2017}, as well as their ease of use \cite{Karras.2017a, Karras.2018b, Karras.2016}, facilitate dissemination, discourse, and feedback via social media platforms such as YouTube \cite{Schneider.2017, Vistisen.2017, Schneider.2019a}. The use of vision videos on social media platforms offers the potential to gather millions of views and solicit thousands of comments \cite{Vistisen.2017}. So far, Schneider and Bertolli \cite{Schneider.2019a} explored the production of vision videos for CrowdRE. They also mentioned the potentially beneficial use of social media platforms to disseminate vision videos for soliciting feedback. However, they explicitly state that the use of vision videos for CrowdRE is future work.

In this paper, we focus on this use of vision videos for CrowdRE. We are interested in the potential of using vision videos on social media platforms to solicit feedback in the form of video comments. We ask the following research question:

\begin{mdframed}
	\textbf{RQ:} What is the potential of using vision videos on social media platforms to solicit feedback in the form of video comments for CrowdRE?
\end{mdframed}

In a case study, we investigate this potential from three perspectives: (1) \textit{quantity}, (2) \textit{content}, and (3) \textit{suitability}. All three perspectives are important to assess the potential of using vision videos for CrowdRE \cite{Khan.2019}. First, we must determine how well vision videos can motivate stakeholders to participate in the crowd, i.e., to watch the videos and write comments. Second, we must examine the content of the comments and their relevance to RE. Third, we must investigate how well the comments are suited for the current analysis techniques in CrowdRE. Based on these three perspectives, we contribute the following insights:

\begin{mdframed}
	\begin{enumerate}[leftmargin=0cm]
		\item[] (1) \textit{Quantity}: Vision videos can solicit many of comments from various viewers in a short period of time.
		
		\item[] (2) \textit{Content}: In addition to typical categories of user feedback, video comments also address the topic \textit{safety}, which has not appeared in analyses of user feedback so~far.
		
		\item[] (3) \textit{Suitability}: Despite certain differences in the performance, machine learning algorithms can already classify video comments well in terms of content.
	\end{enumerate}
\end{mdframed}

This paper is structured as follows: Section \ref{sec:background-and-related-work} discusses related work. While section \ref{sec:case-study} describes the case study design, section \ref{sec:results} reports the findings. After presenting threats to validity in section \ref{sec:threats-to-validity}, the findings are discussed in section \ref{sec:discussion}. Section \ref{sec:conclusion} concludes the paper.

\section{Background and Related Work}
\label{sec:background-and-related-work}
One of the four services of CrowdRE is to provide motivational instruments to engage stakeholders to actively participate in a crowd \cite{Groen.2016}. However, this motivation is a major challenge since simply providing a forum is not enough to solicit feedback and stimulate discussions \cite{Stade.2017}. While most CrowdRE approaches focus on the use of forums and app stores, few consider social media platforms \cite{Khan.2019, Wang.2019, Lim.2021}. Khan et al.~\cite{Khan.2019} argue that social media platforms, with their entertaining and enjoyable activities, e.g., watching a video, provide a great opportunity to motivate stakeholders to participate in a crowd.

Vision videos are one specific kind of videos. These videos are frequently used to solicit feedback and stimulate discussion in RE practices such as workshops and focus groups \cite{Karras.2020d, Karras.2018b, Broll.2007}. Although different researchers \cite{Bennaceur.2016, Darby.2018, Schneider.2019, Busch.2020} reported on the benefits of vision videos for soliciting feedback and stimulating discussions in group or individual meetings, they seldom provide details about the \textit{quantity} and \textit{content} of the feedback.
Bennaceur et al. \cite{Bennaceur.2016} used vision videos in group discussions to solicit as much feedback from the participants as possible. Darby et al. \cite{Darby.2018} produced a vision video to gather feedback on a visionary scenario from a group of stakeholders. While both papers present insights into the main themes of the overall feedback, they do not provide details on the \textit{quantity} or \textit{content} of the individual contributions in the feedback. Schneider et al.~\cite{Schneider.2019} as well as Busch et al. \cite{Busch.2020} used vision videos in individual meetings with single stakeholders to solicit feedback. Both papers examined the number of contributions in the feedback. Although both papers defined what constitutes a contribution, including ideas, questions, requirements and rationale, neither paper explored the \textit{content} of the feedback in detail.
This lack of detail about feedback may be a problem of its direct nature in meetings. In this context, it is far more difficult to grasp, capture, and classify feedback immediately~\cite{Karras.2016a}.

Based on the benefits of vision videos for soliciting feedback and stimulating discussion in meetings, the question arises whether these benefits can also be transferred to social media platforms. On these platforms, (video) comments represent the main communication channel. These comments are similar to app reviews, forum posts, or tweets and thus could be another source of feedback for CrowdRE. Following this line of thought, Vistisen and Poulsen \cite{Vistisen.2017} manually examined 300 comments of a vision video in a case study. Their results indicate that the use of vision videos on social media platforms can solicit valuable feedback from viewers. However, a manual analysis is not feasible in the long term due to the potentially large number of comments. Thus, automated analyses are unavoidable for which video comments must be \textit{suitable} \cite{Groen.2018, Groen.2015}.

Several researchers already applied various automated methods on video comments to analyze and classify them for a wide variety of purposes. Below, we can only give a brief overview based on some approaches due to the large number of use cases in various research areas. There are two main use cases for analyzing video comments.

The first use case is \textit{spam} detection. In this case, \textit{spam} refers to any irrelevant, promotional, incoherent, offensive, or harmful comment \cite{Sharmin.2017, Abdullah.2018, Obadimu.2019, Das.2020}. These comments need to be identified and removed from the social media platforms. 
The second use case is \textit{ham} detection. In this case, \textit{ham} refers to any relevant comment that provides an opinion, question, concern, need or other relevant information from the viewers to the video producers \cite{Asghar.2015, Khan.2016, Poche.2017, Benkhelifa.2018}.
In all cases, the researchers used and compared several algorithms such as \textit{support vector machines}, \textit{naive Bayes}, \textit{logistic regression}, or \textit{neural networks} to classify the video comments for their respective purpose. In accordance with each other, these researchers emphasize the potential of video comments and their automated analysis.

These insights motivated us to further explore the potential of using vision videos on social media platforms for CrowdRE. For this purpose, we carried out a case study.

\section{Case Study}
\label{sec:case-study}
The design and reporting follow the guideline for case study research in software engineering by Runeson and Höst \cite{Runeson.2009}.

\subsection{The Case and Unit of Analysis}
The case is represented by \textit{The Boring Company}\footnote{\url{https://www.boringcompany.com/}}. This company has the mission to solve traffic issues by creating safe, fast-to-build, and cost-effective transportation, utility, and freight tunnels. On April 28, 2017, \textit{The Boring Company} released an animated video\footnote{\url{https://www.youtube.com/watch?v=u5V_VzRrSBI}} on their vision to solve the traffic problems in cities by building large networks of tunnels underneath them. This vision video is the unit of analysis. The video shows a car that uses a network of tunnels underneath a city to travel from one location to another (see \figurename{ \ref{fig:vision_video}).

\begin{figure}[htbp]
	\captionsetup{justification=justified}
	\centering
	\includegraphics[width=0.95\linewidth]{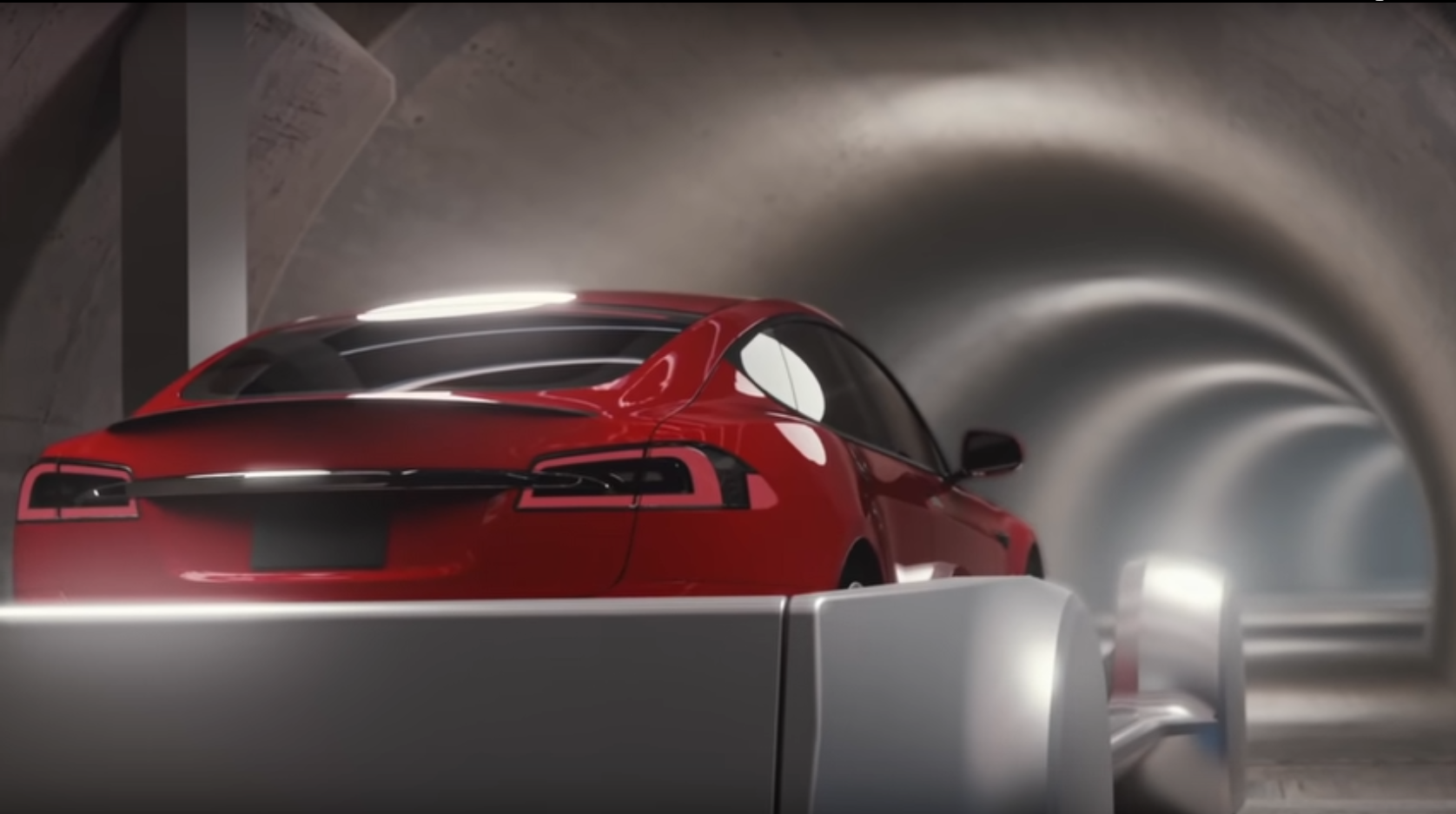}
	\caption{Snapshot from the vision video of \textit{The Boring Company}}
	\label{fig:vision_video}
\end{figure}

\subsection{Data Analysis Procedure}
In \figurename{ \ref{fig:process}}, we present the entire data analysis procedure consisting of four phases: Data collection, data cleaning, manual analysis, and automated analysis.

\begin{figure*}[!t]
	\captionsetup{justification=justified}
	\centering
	\includegraphics[width=0.8\linewidth]{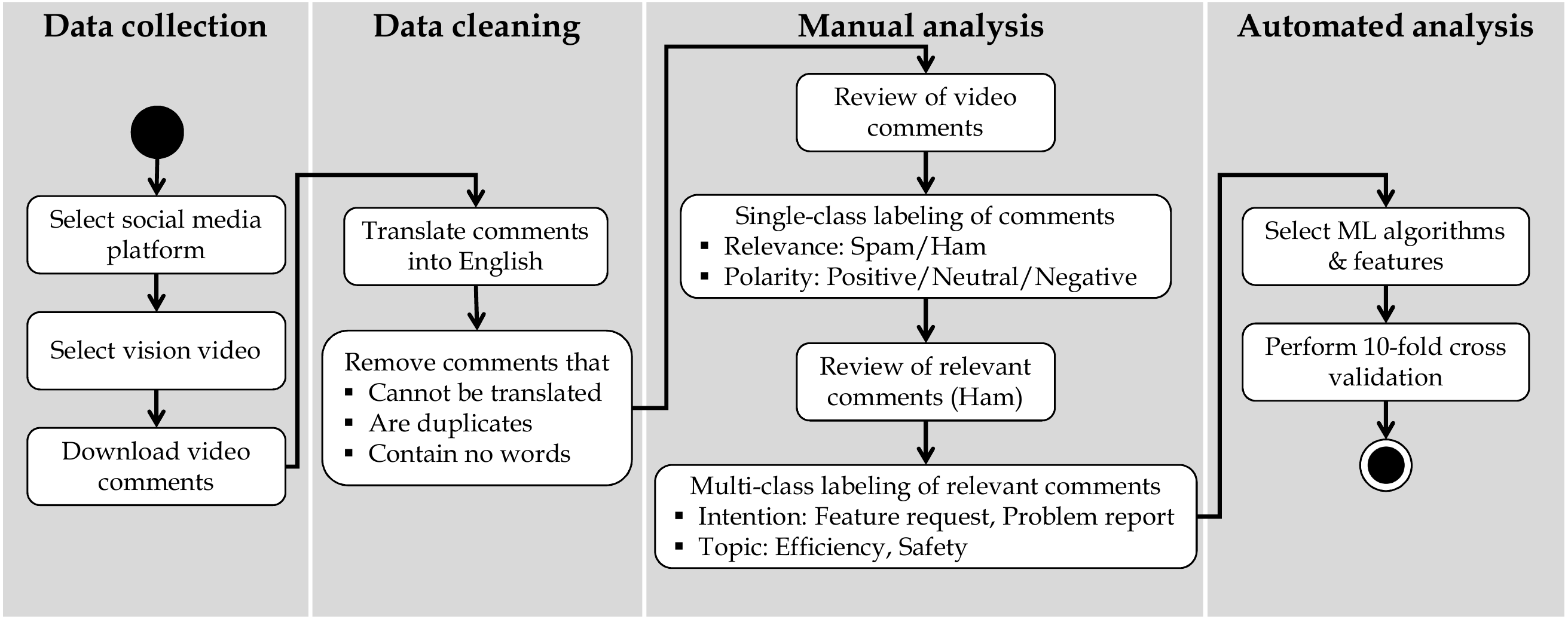}
	\caption{Overview of the data analysis procedure}
	\label{fig:process}
\end{figure*}

\subsubsection{Data Collection}
As social media platform, we selected YouTube since it is the most popular and widely used video-sharing platform where any user can upload, watch, comment, like, dislike, and discuss any videos~\cite{Burgess.2018, Schwemmer.2018, Poche.2017}. Based on a manual search, we found the vision video shared by \textit{The Boring Company} via its channel with 140000 subscribers.  We chose this video due to its large number of comments. On October 13, 2020, we downloaded the data set of the video via the YouTube Data API\footnote{\url{https://developers.google.com/youtube/v3}}. \tablename{~\ref{tbl:video_statistics}} summarizes the statistics of the data set compared to our last access on the video exactly four years after its release. The data set contains 4505 video comments from 4302 different authors. In addition, there are more than 2000 replies that represent discussion threads below specific comments. In this paper, we only focus on the comments since these are directly related to the video. An analysis of the replies remains future work.

\begin{table}[htbp]
	\renewcommand{\arraystretch}{1.3}
	\captionsetup{justification=justified}
	\centering
	\caption{Overview of the statistics of the vision video}
	\label{tbl:video_statistics}
	\resizebox{\columnwidth}{!}{
		\begin{tabular}{|c|c|c|c|c|c|}
			\hline
			\textbf{State} & \textbf{Views} & \textbf{Likes} & \textbf{Dislikes} & \textbf{Comments} & \textbf{Authors} \\ \hline \hline
			\begin{tabular}[c]{@{}c@{}}Oct 13th, 2020\\ (download)\end{tabular} & 6.9 million & 59621 & 4287 & 4505 & 4302 \\ \hline
			\begin{tabular}[c]{@{}c@{}}Apr 28th, 2021\\ (last access)\end{tabular} & 8.3 million & 88457 & 4913 & 5184 & 4980 \\ \hline
	\end{tabular}}
\end{table}

\subsubsection{Data Cleaning}
Although the majority of the comments is written in English, there are several comments written in different languages. We decided to translate these comments into English to keep as many comments in the data set as possible. For this purpose, we manually translated each comment using \textit{Google Translate}\footnote{\url{https://translate.google.com/}}. For data cleaning, we removed all comments from the data set that could not be translated, were exact duplicates, or did not contain any words. The final data set contains 4400 comments for manual analysis.

\subsubsection{Manual Analysis}
The manual analysis served two purposes. First, this qualitative analysis allowed us to explore the potential of using vision videos for CrowdRE from the two perspectives \textit{quantity} and \textit{content}. Second, we had to create a labeled data set for the automated analysis to investigate the third perspective \textit{suitability}. We started with a manual review of the video comments to get an impression of their contents. 

Based on the insights, we performed two classifications. For each classification, we provided the coders with a guide sheet that describes the coding task, defines the categories, and lists examples to reduce inconsistencies and increase the quality of manual labeling \cite{Maalej.2015}. In the first classification, we distinguished between comments that are relevant (\textit{ham}) to RE and those that are irrelevant (\textit{spam}) \cite{Santos.2019}. For this purpose, two coders with a computer science background classified all 4400 video comments regarding their relevance and, in addition, their polarity. This classification was a single-class labeling since each comment can only be \textit{ham} or \textit{spam} and has one polarity. After each coder classified all comments, we organized a virtual meeting to resolve all disagreements (863 regarding relevance and 1869 regarding polarity) with the two~coders. In particular, we discussed each disagreement in the group to reach a consensus between the two coders.

Afterwards, we reviewed all 764 relevant (\textit{ham}) comments to understand their intentions and topics~\cite{Santos.2019}. We found the two main intentions \textit{feature request} and \textit{problem report} as well as the two main topics \textit{efficiency} and \textit{safety}. In the second classification, the same two coders classified all relevant comments regarding these two intentions and two topics. This classification was a multi-class labeling since these categories \enquote{\textit{are not mutually exclusive and can support each other when used in combination}}~\cite[p.~8]{Santos.2019}. After each coder classified all comments, we organized again a virtual meeting to discuss and resolve all disagreements (191 regarding \textit{feature request}, 378 regarding \textit{problem report}, 246 regarding \textit{efficiency}, and 158 regarding \textit{safety}) in the group. We present the detailed results of the manual analysis in the sections \ref{sec:quantity} and \ref{sec:content}.

\subsubsection{Automated Analysis}
\label{sec:automated-analysis}
The automated analysis served to examine the \textit{suitability} of video comments for current analysis techniques in CrowdRE. For this purpose, we decided to follow the overview of user feedback classification approaches \cite{Santos.2019a}.

Based on Santos et al. \cite{Santos.2019a}, we selected the three machine learning (ML) algorithms \textit{support vector machines} (\textit{SVM}), \textit{naive Bayes}, and \textit{random forest}. While the first two algorithms are most frequently applied in CrowdRE, the third one is rarely used but showed very good performance~\cite{Santos.2019a}. We combined each of the three algorithms with the two ML features \textit{bag of words} (BOW) and \textit{term frequency - inverse document frequency} (TF-IDF), which are most frequently used in CrowdRE \cite{Santos.2019a}. The BOW feature included the use of only letters, tokenization, stemming, and the removal of stop words. For the implementation, we used \textit{Python} and the library \textit{scikit-learn}\footnote{\url{https://scikit-learn.org/stable/}} with the provided default settings of the particular algorithms. For each of these six combinations of algorithm and feature, we performed a 10-fold cross validation regarding the binary classification of \textit{ham}, \textit{feature request}, \textit{problem report}, \textit{efficiency}, and \textit{safety}. Each 10-fold cross validation was repeated ten times to reduce the data splitting bias. We initially focused on binary classifiers since multi-class classifiers are even more complicated to use efficiently~\cite{Maalej.2015}. We present the detailed results of the automated analysis in section \ref{sec:suitability}.

We published the raw and labeled data set as well as all additional materials, such as the guide sheets and \textit{Python} scripts, online to increase the transparency of our results \cite{Karras.2021}.

\section{Results}
\label{sec:results}
Below, we present the results of analysis grouped by the three perspectives \textit{quantity}, \textit{content}, and \textit{suitability}.

\subsection{Quantity}
\label{sec:quantity}
This perspective focuses on how well the vision video can motivate stakeholders to participate in the crowd by watching the video and writing comments. As shown in \tablename{ \ref{tbl:video_statistics}}, the video had 6.9 million views and 4505 comments from 4302~different authors at the time we downloaded the data set (after almost 3.5 years). Over the next six months, the number of views, comments, and authors further increased. Although 4~years have passed after its release (see \tablename{ \ref{tbl:video_statistics}}), the video generated another 1.4 million views and 679 comments from 678 different authors. In CrowdRE, it is important to receive a large amount of feedback from many different stakeholders in a short period of time \cite{Groen.2016}. In \figurename{~\ref{fig:comments_authors}}, we illustrate the distribution of the number of comments and different authors over time. In the first four days after its release, the vision video solicited a total number of 2770 comments from 2660 different authors, which is 53.4\% of all comments and 53.4\% of all authors in four years. In the following 3 years, the number of comments increased slightly by 1142 (22.1\%). In the fourth year, the video solicited 1272 comments (24.5\%). Based on data set, we cannot explain this sharp increase in the last year. 

\begin{mdframed}
	\textlabel{\protect Finding 1}{f1}\textbf{Finding 1:}
	The vision video motivated viewers to participate in the crowd. In a short period of time (four~days), a large number of viewers watched the video and 2660 of them provided feedback by writing at least one comment.
\end{mdframed}

\begin{figure}[htbp]
	\captionsetup{justification=justified}
	\centering
	\includegraphics[width=0.95\linewidth]{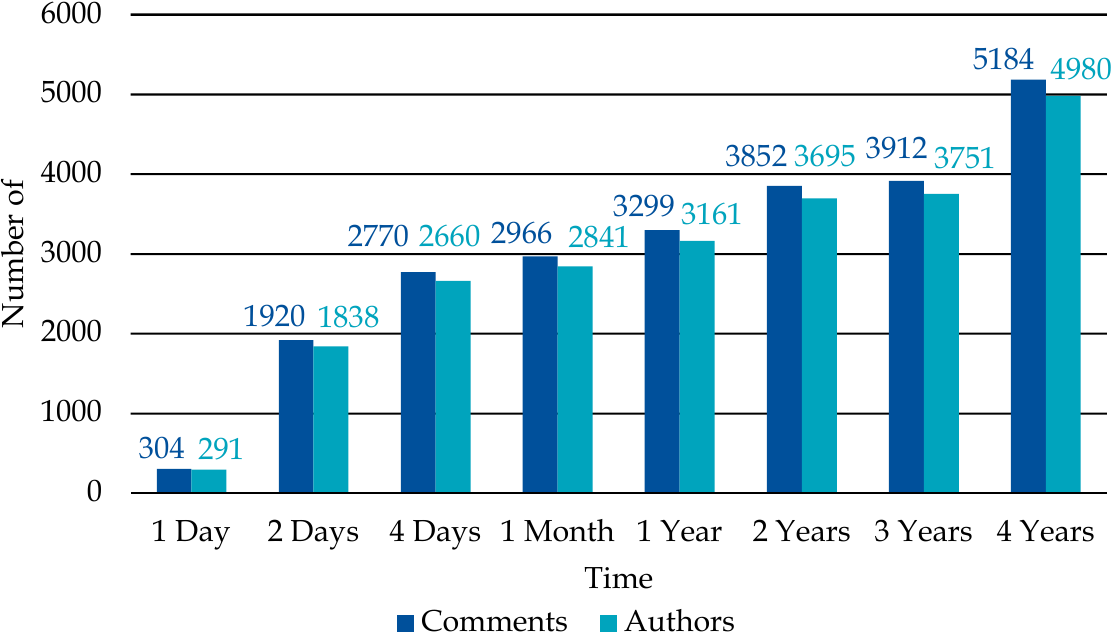}
	\caption{Number of comments and different authors over time}
	\label{fig:comments_authors}
\end{figure}

\subsection{Content}
\label{sec:content}
This perspective considers the content of the comments and their relevance to RE. We summarized the results of the single- and multi-class labeling in \tablename{ \ref{tbl:labeling}}. Based on the single-class labeling, we identified 764 of the 4400 comments (17\%) as relevant (\textit{ham}) for RE. These comments include the authors' opinions, questions, concerns, needs, and ideas combined with constructive feedback about the presented vision. The single-class labeling was restrictive since we determined that any comment that is promotional, incoherent, offensive, harmful, or lacked constructive feedback should be labeled as irrelevant (\textit{spam}). As a result, the proportion of relevant comments is rather small. For example, the comment 3914 \enquote{\textit{This Is Awesome.}} is a positive statement on the vision but lacks constructive feedback. Thus, the comment was classified as \textit{spam}. In contrast, one author wrote the comment 517 \enquote{\textit{Good idea but what about the possibility of sinkholes}}. The comment was classified as \textit{ham} since it provides constructive feedback by pointing out a potential problem.

\newcommand*{\MyIndent}{\hspace*{0.25cm}}%
\begin{table}[htbp]
	\renewcommand{\arraystretch}{1.3}
	\captionsetup{justification=justified}
	\centering
	\caption{Results of the labeling of the video comments}
	\label{tbl:labeling}
	\begin{tabular}{|l|c|c|c|c|}
		\hline
		\multicolumn{1}{|c|}{\textbf{Category}} & \textbf{Positive} & \textbf{Neutral} & \textbf{Negative} & \textbf{Overall} \\ \hline \hline
		\textbf{Overall} & 400 & 3033 & 967 & 4400 \\ \hline
		\textbf{Spam} & 368 & 2486 & 782 & 3636 \\ \hline
		\textbf{Ham} & 32 & 547 & 185 & 764 \\ \cline{2-5}
		\MyIndent \textbf{Feature request} & 7 & 115 & 20 & 142 \\ \cline{2-5}
		\MyIndent \textbf{Problem report} & 16 & 277 & 116 & 409 \\ \cline{2-5}
		\MyIndent \textbf{Efficiency} & 5 & 101 & 54 & 160 \\ \cline{2-5}
		\MyIndent \textbf{Safety} & 16 & 205 & 91 & 312 \\ \hline
	\end{tabular}
\end{table}

We further investigated the content of the relevant comments manually. In this way, we achieved a better understanding of the intentions of the comments. We found all three intentions \textit{informing}, \textit{reporting}, and \textit{requesting} corresponding to the taxonomy for user feedback classification by Santos et al.~\cite{Santos.2019}. \textit{Informing} comments provide reasons based on features or problems either for or against the vision presented in the video, e.g., comment 151 \enquote{\textit{A metro rail with extra steps, inefficiency, and fuel cost.}} \textit{Reporting} comments address problems of the vision suspected by the author. One author wrote comment~271 \enquote{\textit{Very unsafe I see no emergency exit at all}}. \textit{Requesting} comments focus on features of the vision. These comments either ask for missing functionality or share ideas for improving the vision by adding or changing features. For example, comment 302 \enquote{\textit{it would be insane if you could charge your car while riding the tunnel}}. We simplified the multi-class labeling by initially focusing only on \textit{feature request} and \textit{problem report} as the two main intentions, similar to Maalej and Nabil \cite{Maalej.2015}.

In addition to the manual review, we used word clouds to explore frequent words and thus main topics in comments~\cite{Obadimu.2019, Heimerl.2014}. In this way, we found that the two topics \textit{efficiency} (see comment 151) and \textit{safety} (see comment 271) are mainly addressed by the authors. These two intentions and these two topics form the categories for the multi-class labeling. For example, the following comment~1289 covers both intentions and both topics: \enquote{\textit{As mentioned on other social media it may be a good idea to look into muti level above ground tunnels} [(feature request)]\textit{, although affected more by the weather eliments} [(problem report)] \textit{would be easier to maintain} [(efficiency)] \textit{and possibly much safer in the event of an earthquake or flooding} [(safety)]}.

\begin{mdframed}
	\textlabel{\protect Finding 2}{f2}\textbf{Finding 2:}
	The content of the comments is related to taxonomy of user feedback classification categories and thus relevant to RE. The relevant comments are mainly related to \textit{feature request}s or \textit{problem report}s regarding the presented vision. These comments primarily address the topics \textit{efficiency} and \textit{safety} of the envisioned system.
\end{mdframed}


\subsection{Suitability}
\label{sec:suitability}
This perspective investigates how well video comments are suitable for the current analysis techniques in CrowdRE. For this purpose, we performed an automated analysis of the video comments as described in section \ref{sec:automated-analysis}. In consideration of the highly varying numbers of comments in each category (see~\tablename{ \ref{tbl:labeling}}), we decided to initially analyze a balanced data set for each category. Each balanced data set consists of as many data points as possible, e.g., the balanced data set on \textit{safety} contains 312 comments on \textit{safety} and 312 others randomly selected. \tablename{ \ref{tbl:performance}} presents the average performance of the algorithms grouped by ML features and categories.

\begin{table}[htbp]
	\renewcommand{\arraystretch}{1.3}
	\captionsetup{justification=justified}
	\centering
	\caption{Average performance of the algorithms based on a 10 times repeated 10-fold cross validation\protect\footnotemark}
	\label{tbl:performance}
	\includegraphics[width=0.95\linewidth]{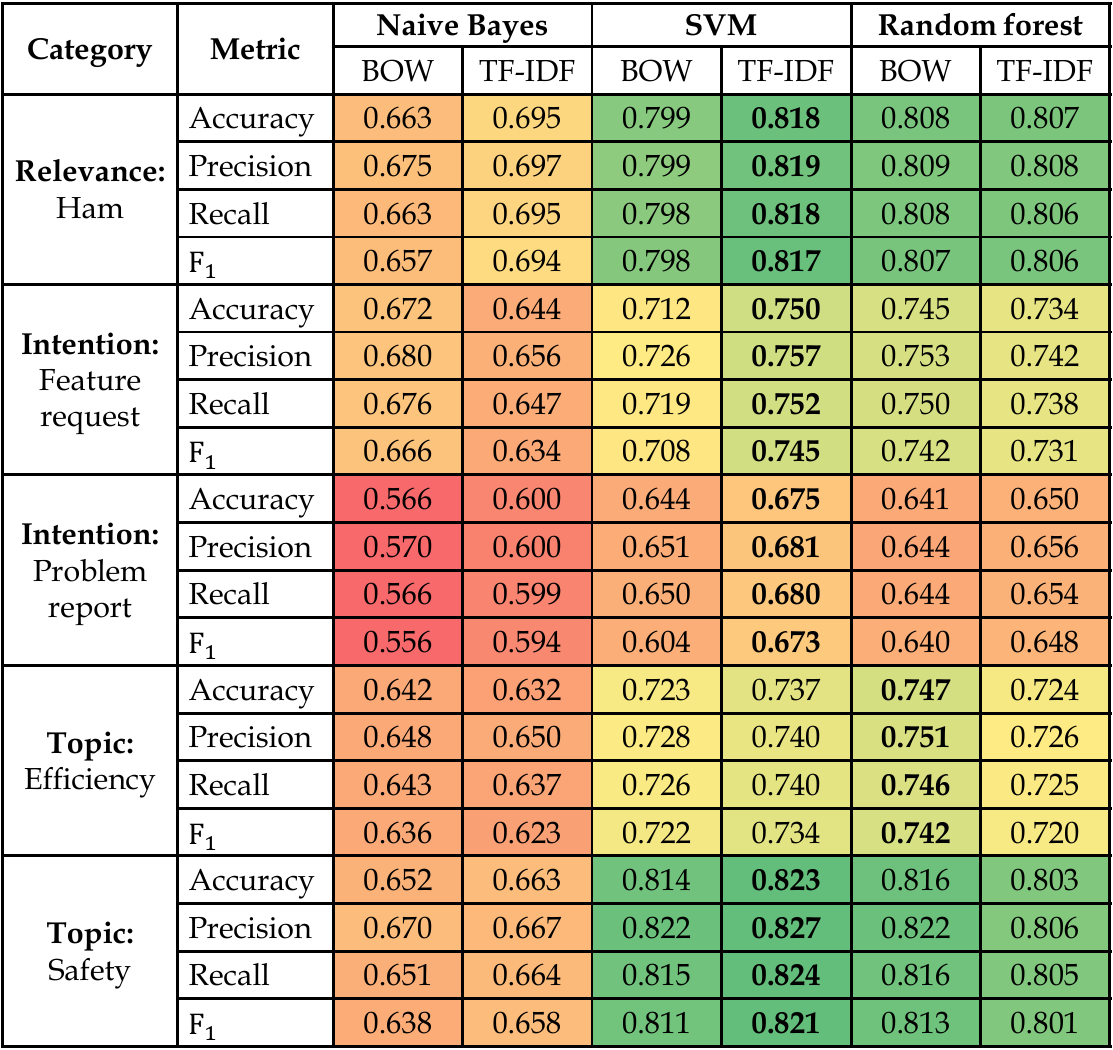}
\end{table}

\footnotetext{The bold numbers highlight the best values per row. The conditional formatting is based on a comparison of all values per metric and the color gradient is from green (highest value) to yellow to red (lowest value).}

The analysis of all five categories shows promising results since the $F_{1}$ values range from 0.556 up to 0.821. However, the results obviously differ among the categories and algorithms.

\subsubsection{Categories}
While the relevant (\textit{ham}) comments can be classified well by all algorithms (0.657 $< F_{1} <$ 0.817), the performance of the other classifications regarding the intentions and topics is divergent. The intention \textit{problem report} consistently shows the lowest $F_{1}$ values (below 0.673) indicating difficulties in its classification. The other intention \textit{feature request} and the topic \textit{efficiency} have moderate results (0.623 $< F_{1} <$ 0.745). The topic \textit{safety} is one of the best to classify. In case of using \textit{SVM} with TF-IDF, we achieved the highest $F_{1}$ value of 0.821.

\subsubsection{Algorithms}
\textit{Naive Bayes} consistently shows the lowest results for all categories (0.556 $< F_{1} <$ 0.666). The results of \textit{SVM} (0.604 $< F_{1} <$ 0.821) and \textit{random forest} (0.640 $< F_{1} <$ 0.813) are always better than those of \textit{naive Bayes} for each category. \textit{SVM} and \textit{random forest}, in turn, achieve very similar results. Nevertheless, in almost all cases, \textit{SVM} with TF-IDF achieves slightly better results than \textit{random forest} (see \tablename{~\ref{tbl:performance}, bold highlighting}). Only for the topic \textit{efficiency} are the results of \textit{random forest} with BOW slightly better.

These observations indicate that video comments are suitalbe for current analysis techniques in CrowdRE. Already the use of algorithms with their default settings (see section \ref{sec:automated-analysis}) leads to first promising results. Even though the algorithms produce different results, the use of \textit{SVM} with TI-IDF shows the best results overall.

\begin{mdframed}
	\textlabel{\protect Finding 3}{f3}\textbf{Finding 3:}
	The video comments are suitable for current analysis techniques in CrowdRE. Despite differences in the performance of the algorithms used and the individual categories, the automated analysis shows promising results. While \textit{problem report}s are more difficult to classify, especially relevant (\textit{ham}) comments and comments on \textit{safety} can be classified well. In particular, the use of \textit{SVM} with TF-IDF achieved the best results overall.
\end{mdframed}

\section{Threats to Validity}
\label{sec:threats-to-validity}
Below, we report threats to construct, external, internal, and conclusion validity according to Wohlin et al. \cite{Wohlin.2012}.

\subsection{Construct Validity}
There is a mono-operation bias since the study is based on one case. Although the case study is a real-world example, one case cannot represent the complexity in reality. In addition, we must emphasize that this case is special since the vision video comes from a company founded by Elon Musk and already has a strong social media presence. The selection of three ML algorithms and two features leads to a method bias (see section \ref{sec:automated-analysis}). According to Santos et al. \cite{Santos.2019a}, there are at least six other algorithms and five other features that are used in CrowdRE. We assume that our selection is a good compromise for an initial study since we focus on the two most frequently used algorithms and features. In addition, we use an algorithm that is rarely used but has shown a very good performance \cite{Santos.2019a}. However, the selection is necessary since this study first investigates whether the use of vision videos on social media platforms has any potential to solicit feedback in the form of video comments for CrowdRE. For this reason, we also neglected the tuning of the hyperparameters and used the algorithms with their default settings. As a result, this case study is only a preliminary investigation and future replications and extensions are needed to strengthen the results and draw more reliable conclusions. 

\subsection{External Validity}
The single case limits the generalization of the results. However, generalization is not the focus of this study. Instead, we want to get a first impression of the potential of using vision videos on social media platforms for CrowdRE. First, we attempted to find existing labeled data sets that we could have used to increase external validity. However, all of the data sets we found only contained labeled video comments on coding tutorials, music videos, or movie trailers and thus did not fit our context and purpose. We also contacted Vistisen and Poulsen \cite{Vistisen.2017} to obtain their data set. Unfortunately, they only created a data set that contains the raw data without the labels. For future extensions of our case study, we currently create additional data sets based on the raw data set of Vistisen and Poulsen \cite{Vistisen.2017} as well as another vision video\footnote{\url{https://www.youtube.com/watch?v=S5fOWB6SNqs}} from YouTube.

\subsection{Internal Validity}
The video comments often have a poor text quality and belong to several categories, even beyond those categories classified in this study, such as \textit{user experience} and \textit{ratings} found by Maalej and Nabil \cite{Maalej.2015}. These shortcomings impede the manual single- and multi-class labeling. We provided the coders with a guide sheet for their orientation to mitigate this threat to validity \cite{Neuendorf.2015}. After each labeling, we calculated \textit{Cohen's} $\bm{\kappa}$ \cite{Cohen.1960} for each category (see \tablename{~\ref{tbl:kappa}}). We found a maximum value of 0.35, indicating only moderate agreement~\cite{Landis.1977} and thus high subjectivity of the classification by the two coders. For this reason, we organized two meetings to discuss each disagreement in the group and reach a consensus between the two coders. As suggested by Maalej and Nabil~\cite{Maalej.2015}, this procedure helps to resolve disagreements and reach a consensus. Nevertheless, we cannot guarantee that all video comments are completely and correctly classified. Furthermore, we simplified the labeling by focusing only on five categories. According to the taxonomy of user feedback classification categories \cite{Santos.2019}, there are many more (sub-) categories. However, we neglected these categories for now due to the preliminary nature of our study.

\begin{table}[htbp]
	\renewcommand{\arraystretch}{1.3}
	\captionsetup{justification=justified}
	\centering
	\caption{Inter-rater reliability between the two coders}
	\label{tbl:kappa}
	\resizebox{\columnwidth}{!}{
		\begin{tabular}{|l|c|c|c|c|c|c|}
			\hline
			\textbf{Category} & \textbf{Relevance} & \textbf{Polarity} & \textbf{\begin{tabular}[c]{@{}c@{}}Feature\\ request\end{tabular}} & \textbf{\begin{tabular}[c]{@{}c@{}}Problem\\ report\end{tabular}} & \textbf{Efficiency} & \textbf{Safety} \\ \hline \hline
			\textbf{Cohen's $\bm{\kappa}$} & 0.35 & 0.14 & 0.15 & -0.01 & 0.28 & 0.26 \\ \hline
		\end{tabular}
	}
\end{table}

\subsection{Conclusion Validity}
All our findings are based on the results of the manual analysis and thus the subjective interpretation of the video comments. We cannot completely exclude the misinterpretation of some video comments due to a missing context. Nevertheless, we mitigated this threat to validity with a defined strategy. While the first two authors planned the data analysis procedure, the third author reviewed the design and suggested improvements. Afterwards, the second author performed the entire data analysis procedure with two external coders, the first author supervised the procedure, and the third author reviewed the analysis and results \cite{Kristo.2021a}. For this reason, we are confident that the findings are valid and sound.

\section{Discussion}
\label{sec:discussion}
The case study provides important insights into the potential of using vision videos on social media platforms to solicit feedback in form of video comments for CrowdRE. Our findings substantiate that vision videos (1) can solicit many comments from various viewers in a short period of time, which (2) contain relevant content to RE, and (3) are suitable for current analysis techniques in CrowdRE.

\textit{The use of vision videos on social media platforms shows promise for soliciting feedback in form of video comments}. We investigated the potential of using vision videos for CrowdRE from the three perspectives \textit{quantity}, \textit{content}, and \textit{suitability}. 
The manual analysis shows that the use of the vision video on YouTube resulted in thousands of views and comments within a few days (see \ref{f1}). Although we found that only a certain fraction of these comments are relevant to RE (see \ref{f2}), these comments address typical intentions and topics of user feedback, such as \textit{feature request}, \textit{problem report}, and \textit{efficiency}. Despite the lower fraction of relevant comments, automated analyses are unavoidable due to the large number of comments. The automated analysis shows that video comments are suitable to be analyzed by ML algorithms (frequently) used in CrowdRE. Already algorithms with their default settings achieved a good performance (see~\ref{f3}). For these reasons, we conclude that using vision videos on social media platforms offers great potential to solicit video comments as a source of feedback for CrowdRE.

Most of our findings correspond with the current state of knowledge in CrowdRE literature. However, we also found one new insight. While the topic \textit{safety} is addressed in over 300 video comments, the taxonomy of user feedback classification categories does not contain this characteristic in the \textit{quality in use} category \cite{Santos.2019}. Our topic \textit{safety} corresponds to the characteristic \textit{freedom of risk} of ISO/IEC FDIS 25010:2010~\cite{ISO25010.2010} addressing economic, health, safety, and environmental risks. We assume that this difference is a result of using a vision video as stimulus showing a cyber-physical system. Previous studies have primarily focused on user feedback on pure software systems leading to a taxonomy for user feedback classification \cite{Santos.2019} that does not include \textit{safety} as a characteristic.
Vision videos can stimulate this kind of feedback since they are concrete by showing specific scenarios and contexts \cite{Karras.2016}. Thus, vision videos do not only show the software system but also its use in concrete situations. Viewers can more easily put themselves in these concrete situations \cite{Ferrari.2015, Liskin.2015, Busch.2020}. In this way, they experience the system before it has even been developed. Through the use of vision videos, we are able to empower stakeholders to report their suspected \textit{quality in use} problems, such as economic, health, safety, and environmental risks, even if they have never actually used the system.

The vision video used in this case study is one important point we need to discuss. The video comes from a company founded by Elon Musk and already has a strong social media presence. As a result, this case is special since the conditions to receive feedback were much better than for average companies. Successfully soliciting feedback in the form of video comments requires a good strategy for building a corresponding social media presence for a company. This strategy must go beyond marketing for bridging the gap to the development so that a cycle can be created in the sense of CrowdRE to solicit, analyze, implement, and validate feedback in order to solicit new feedback again. For this purpose, future research must develop systematic and holistic approaches aligned with such a strategy and covering the entire software development process to operationally involve a crowd, e.g., to provide content, form a social network, and support the development with feedback.

For the analysis of the feedback, we currently used traditional and established machine learning algorithms and features. Even though the results are positive, there is a new trend in text classification using embeddings and transformer-based classifiers such as BERT \cite{Devlin.2018}. First approaches using deep learning algorithms such as BERT show promising results in classifying user feedback \cite{Reddy.2021}. Our results show that video comments are another source of feedback whose analysis needs to be further developed, especially considering this current research trend. As an answer to our research question, we can summarize:

\begin{mdframed}
	\textbf{Answer:} The use of vision videos on social media platforms offers a large potential to solicit feedback in form of video comments for CrowdRE. According to the manual and automated analysis, a sufficient number of comments emerge whose content is relevant to RE and which are suitable for current analysis techniques in CrowdRE. The findings point to the conclusion that using vision videos for CrowdRE can motivate stakeholders to actively participate in a crowd by writing video comments that are a valuable source of feedback.
\end{mdframed}

\section{Conclusion \& Future Work}
\label{sec:conclusion}
In this paper, we present the findings of an initial case study on the potential of using of vision videos for CrowdRE. According to these findings, the use of vision videos for CrowdRE has a large potential since these videos can solicit a large amount of feedback from various stakeholders in a short period of time. In addition to typical user feedback categories, such as \textit{feature request} and \textit{problem report}, we found that vision videos empower viewers to experience the future system in a way that allows them to comment on suspected problems related to \textit{safety}. This finding is particularly interesting since this topic is an entirely new category that has not previously appeared in analyses of user feedback. However, the findings should not be overgeneralized due to the preliminary nature of our case study. Nevertheless, we are optimistic that vision videos can motivate stakeholder to actively participate in a crowd, solicit feedback, and stimulate discussion for CrowdRE. While our case study substantiates the first two points, the stimulation of discussions is only an assumption based on a manual review of the replies. Similar to Vistisen and Poulsen~\cite{Vistisen.2017}, we have the impression that replies do not address the video itself but primarily discuss the associated comment. This assumption needs to be investigated in the future, for example with the \textit{CrowdRE-Arg} framework by Khan et al. \cite{Khan.2020}. This framework can identify arguments that are for or against a given statement. Overall, the findings show a positive trend, but future replications and extensions of this study are needed for a more comprehensive assessment of the potential of using vision videos for CrowdRE.

\bibliographystyle{IEEEtran}
\balance
\bibliography{IEEEabrv,references}

\end{document}